\documentclass[a4paper]{PoS}

\title{Infrared variability of radio-loud narrow-line Seyfert 1 galaxies}

\ShortTitle{Infrared variability of radio-loud NLS1 sources}

\author{\speaker{K. \'E. Gab\'anyi,$^{ab}$}\thanks{K\'EG was supported by the J\'anos Bolyai Research Scholarship of the Hungarian Academy of Sciences. K\'EG and SF thank the Hungarian National Research, Development, and Innovation Office (OTKA NN110333) for support. This publication makes use of data products from the WISE, which is a joint project of the University of California, Los Angeles, and the Jet Propulsion Laboratory/California Institute of Technology, funded by the National Aeronautics and Space Administration.
This conference has been organized with the support of the
Department of Physics and Astronomy ``Galileo Galilei'', the 
University of Padova, the National Institute of Astrophysics 
INAF, the Padova Planetarium, and the RadioNet consortium. 
RadioNet has received funding from the European Union's
Horizon 2020 research and innovation programme under 
grant agreement No~730562. 
In loving memory of Zolt\'an.} A. Mo\'or$^{b}$ and S. Frey$^{b}$\\
        \llap{$^a$}MTA-ELTE Extragalactic Astrophysics Research Group, ELTE TTK, P\'azm\'any P\'eter s\'et\'any 1/A, H-1117 Budapest, Hungary \\
	\llap{$^b$}Konkoly Observatory, MTA Research Centre for Astronomy and Earth Sciences, Konkoly Thege Mikl\'os \'ut 15-17, H-1121 Budapest, Hungary \\
        E-mail: \email{krisztina.g@gmail.com}, \email{moor@konkoly.hu}, \email{frey.sandor@csfk.mta.hu}}

\abstract{Most of the radio-loud narrow-line Seyfert 1 (NLS1) galaxies resemble compact steep-spectrum sources. However, the extremely radio-loud ones show blazar-like characteristics, like flat radio spectra, compact radio cores, substantial variability and high brightness temperatures. These objects are thought to be similar to blazars as they possess relativistic jets seen at small angle to the line of sight. This claim has been further supported by the {\it Fermi} satellite discovery of gamma-ray emission from a handful of these sources. Using the {\it Wide-Field Infrared Survey Explorer} ({\it WISE}) data, we analyzed the mid-infrared variability characteristics of $42$ radio-loud NLS1 at $3.4$ and $4.6\,\mu$m. We found that $27$ out of the studied $42$ sources showed variability in at least one of the two infrared bands. In some cases, significant changes in the infrared colors can alter the location of the source in the {\it WISE} color--color diagram which might lead to different classification. More than $60$\% of the variable sources also showed variability within a $1-1.5$ day interval. Such short time scales argue for a compact emission region like those associated with the jets. This connection is further strengthened by the fact that the brightest $\gamma$-ray emitters of the sample ($6$ sources), all showed short time scale infrared variability.}

\FullConference{Revisiting narrow-line Seyfert 1 galaxies and their place in the Universe - NLS1 Padova\\
		9-13 April 2018 \\
		Padova Botanical Garden, Italy}

\begin{document}

\section{Introduction}
Narrow-line Seyfert 1 galaxies (NLS1), a subclass of active galactic nuclei (AGN) are thought to contain a relatively low-mass central black hole ($10^6-10^8 \mathrm{M}_\odot$, \cite{Mathur2000}) and accrete at high rates \cite{CollinKawaguchi}. The fraction of radio-loud ones among NLS1 sources is $\sim 7$\% similar to that of the whole AGN population \cite{Komossa2006}. The extremely radio-loud ones ($\approx 2.5$\% of the NLS1 sources \cite{Komossa2006}) show blazar-like characteristics, like flat radio spectra, compact radio cores, substantial variability and high brightness temperatures. 
Caccianiga et al. \cite{cacci} studied the mid-infrared characteristics of $42$ radio-loud NLS1 sources with blazar-like characteristics and found that the radio and mid-infrared properties can be explained by the mixture of different components: the relativistic jets, star formation in the host galaxy, and the dusty torus. Using the available {\it Wide-Field Infrared Survey Explorer} ({\it WISE}, \cite{wise}) satellite data, we investigated the variability characteristics of these $42$ sources in the mid-infrared bands. 

\section{Analysis of {\it WISE} data}

The {\it WISE} \cite{wise} satellite scanned the whole sky at four infrared (IR) bands, at $3.4$, $4.6$, $12$, and $22\,\mu$m (referred as W1, W2, W3, and W4) during 2010. After the end of the WISE mission, the survey has continued in the framework of the NEOWISE (Near-Earth Object WISE) project \cite{neowise} after September 2010 first for four months. After $34$ months of hibernation of the satellite, the NEOWISE  Reactivation Mission continued. Along its mission, {\it WISE} observed the same regions of the sky  every $\sim 180$ days. Therefore, during the original mission, each source was observed in two mission phases. After the approximately 3-year hibernation gap, another $5$ to $7$ epochs are added in the NEOWISE mission. However, because of the depleted cooling material, the W3 and W4 detectors could not be used any more. NEOWISE only makes measurements at the two shortest wavelength bands (W1 and W2). Moreover, not all sources have the same coverage. Due to the scanning strategy, the number of frames covering a given source depends on its ecliptic coordinates. The closer an object to the ecliptic poles, the more frames of it are obtained. Therefore, the numbers of frames within each mission phase are different for different objects.

\begin{figure}
\begin{minipage}[t]{0.45\textwidth}
\centering
\includegraphics[bb=0 0 550 480, width=0.95\textwidth,clip]{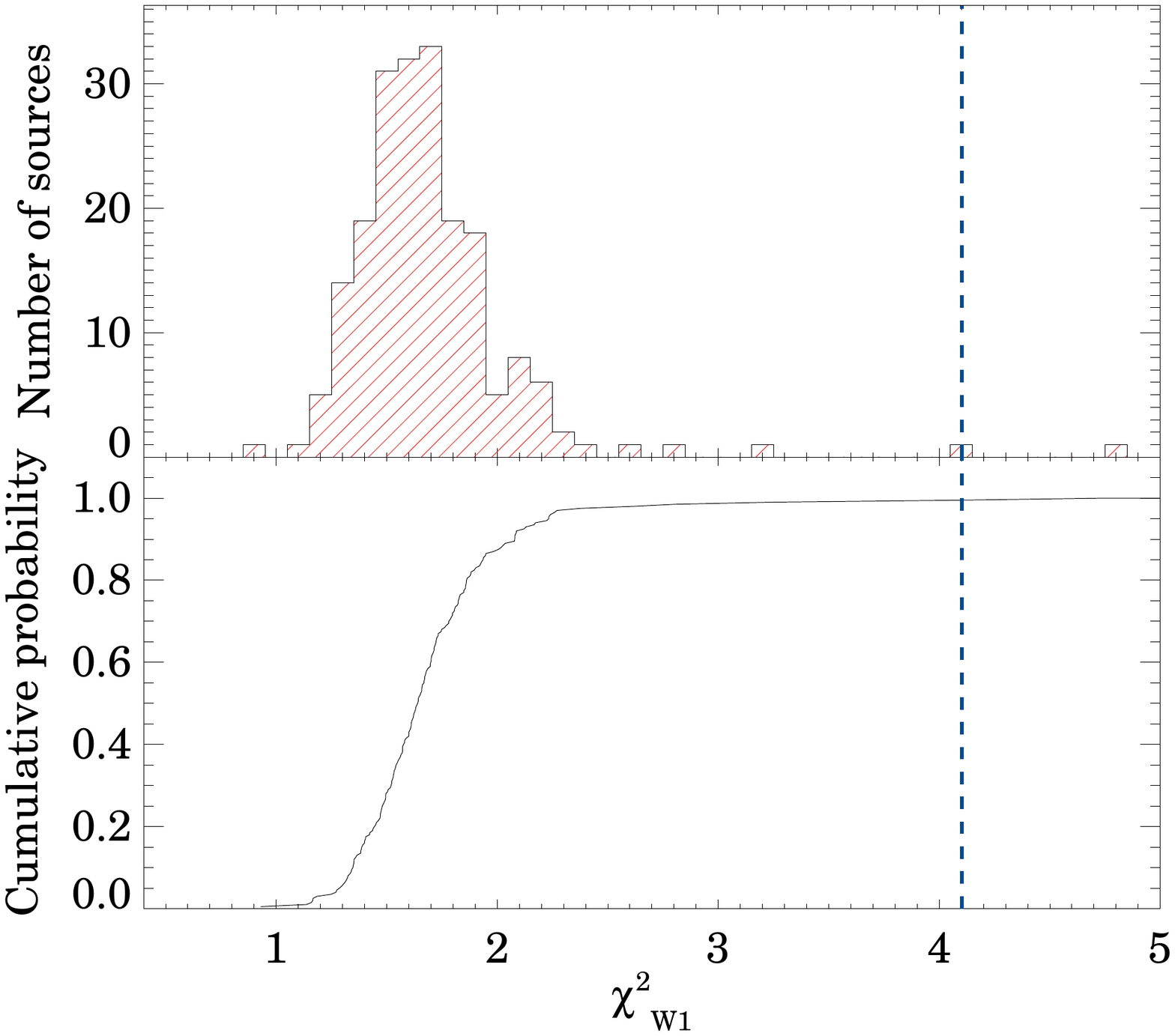}
\caption{Histogram (top) and cumulative distribution (bottom) of $\chi^2_\mathrm{r}$ in W1 band of the nearest $200$ sources around J1238$+$3942. Dashed blue lines indicate the $\chi^2_\mathrm{r}$ value of J1238$+$3942.}
\label{fig:histo}
\end{minipage}
\hfill
\begin{minipage}[t]{0.52\textwidth}
\centering
\includegraphics[bb=10 5 780 565, width=1.05\textwidth,clip]{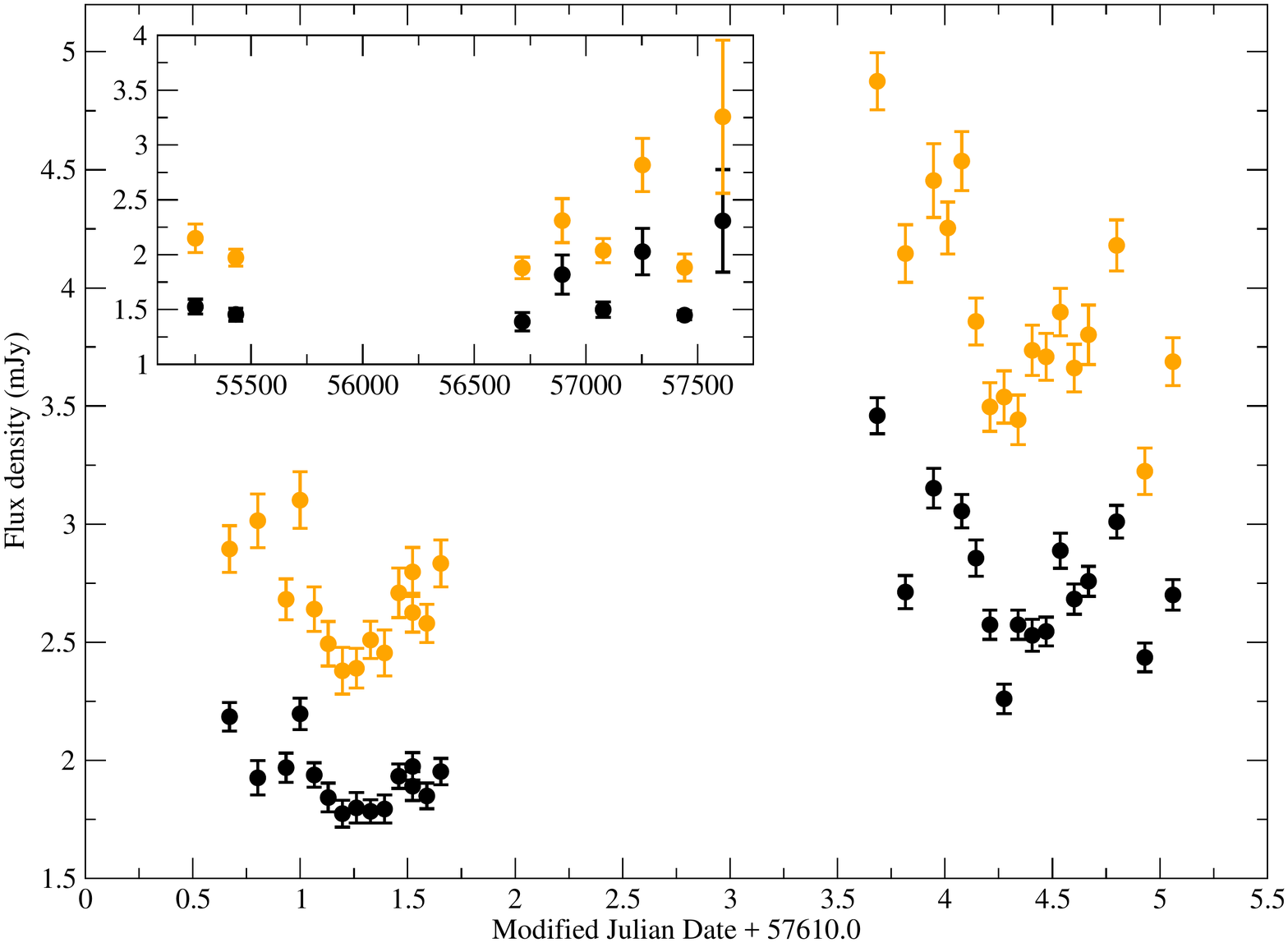}
\caption{Infrared light curve of J1644$+$2619 from the last analysed mission phase. Black dots are for band W1, yellow dots are for band W2. The inset shows the {\it WISE} flux density measurements in these two bands averaged for each of the $8$ mission phases; the error bars represent the variability within the given phases.}
\label{fig:J1644}
\end{minipage}
\end{figure}

We downloaded up to 2016 the {\it WISE} single exposure data\footnote{{http://irsa.ipac.caltech.edu/Missions/wise.html}} of the $42$ radio-loud NLS1 sources from Caccianiga et al. \cite{cacci}. We discarded bad quality data 
following the descriptions in the Exploratory Supplement Series\footnote{{http://wise2.ipac.caltech.edu/docs/release/allwise/expsup/sec3\_2.html}}. We obtained light curves in the W1 and W2 bands for all sources. For most of the sources, there are light curves from $8$ mission phases. For $4$ and $6$ sources, there are light curves from $7$ and $9$ mission phases, respectively. Within each mission phase, the individual measurements of the sources provide short-term light curves with a length $\approx 1$ day, and with measurements in every $\sim 1.5 - 2$\,h. In several mission phases, $2-6$ days long gaps occur as well (see an example in Fig. \ref{fig:J1644}). In the other two bands, most of the sources were observed only in the first mission phase, $8$ sources have measurements in the W3 and W4 from the first two mission phases. In general the signal-to-noise ratios are lower and the sources are fainter in W3 and especially in W4. There were $1$ and $6$ sources with less than three measurement points in bands W3 and W4, respectively.

To determine the variability characteristics of the sources in W1 and W2, we calculated the Stetson index \cite{Stetson} and after converting the magnitudes to flux densities \cite{wise} the reduced $\chi^2$ ($\chi^2_\mathrm{r}$) values. We considered a source variable if its Stetson index is larger than $1$. In all such sources the calculated $\chi^2_\mathrm{r}$ values in both bands were larger than $3$, which also strongly supports the variability claim in these cases.
The Stetson index is more sensitive to the correlated variations measured simultaneously at two bands. Therefore, we also critically examined those objects for which the Stetson index was lower, between $0.4$ and $1$. In all of these objects the $\chi^2_\mathrm{r}$ values exceeded $3$ in at least one of the bands. We then downloaded {\it WISE} data of the nearest $200$ objects with similar brightness in W1 and W2 and calculated their $\chi^2_\mathrm{r}$ values. Based upon the obtained histograms and cumulative distributions (see an example in Fig. \ref{fig:histo}), we deemed a source variable if its $\chi^2_\mathrm{r}$ in a given band was larger than $95$\% of those of the neighbouring sources.

To characterise the strength of variability, we calculated the modulation indices for the variable sources. The modulation index is defined as $m=100 \cdot \sigma/ \langle S\rangle$, where $\sigma$ is the standard deviation of the measured flux density and $\langle S \rangle$ is the average flux density. To study short time scale variability, we also calculated $\chi^2_\mathrm{r}$ for each mission phase separately. 

Due to the lower signal-to-noise ratios and shorter time baselines, we did not perform such detailed variability analysis in bands W3 and W4. We calculated the $\chi^2_\mathrm{r}$ for sources having more measurement points than $3$.

\section{Results}

\begin{figure}
\begin{minipage}[t]{0.5\textwidth}
\centering
\includegraphics[bb=45 55 715 530,width=\textwidth,clip]{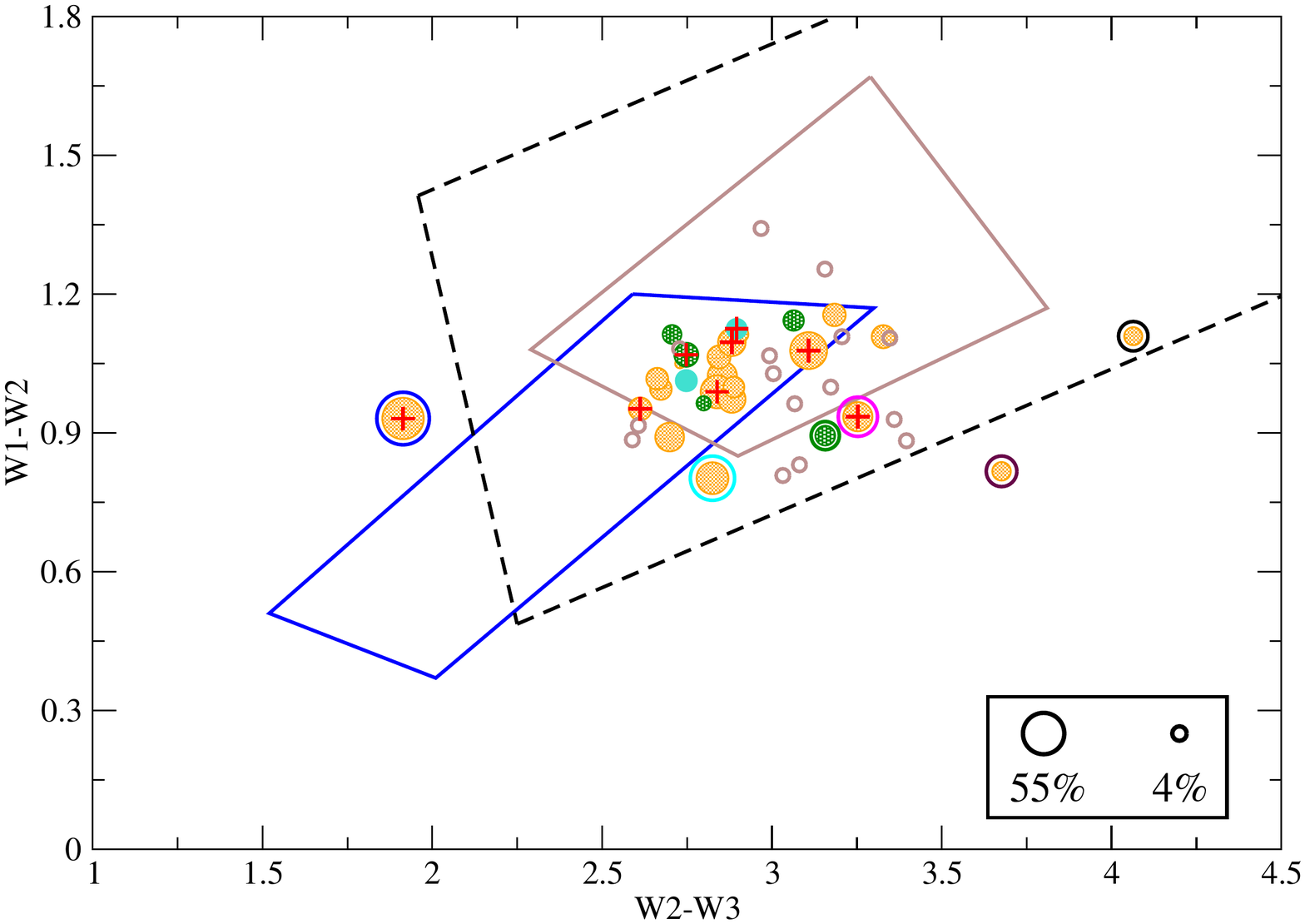}
\end{minipage}
\begin{minipage}[t]{0.5\textwidth}
\centering
\includegraphics[bb=45 35 720 530,width=\textwidth,clip]{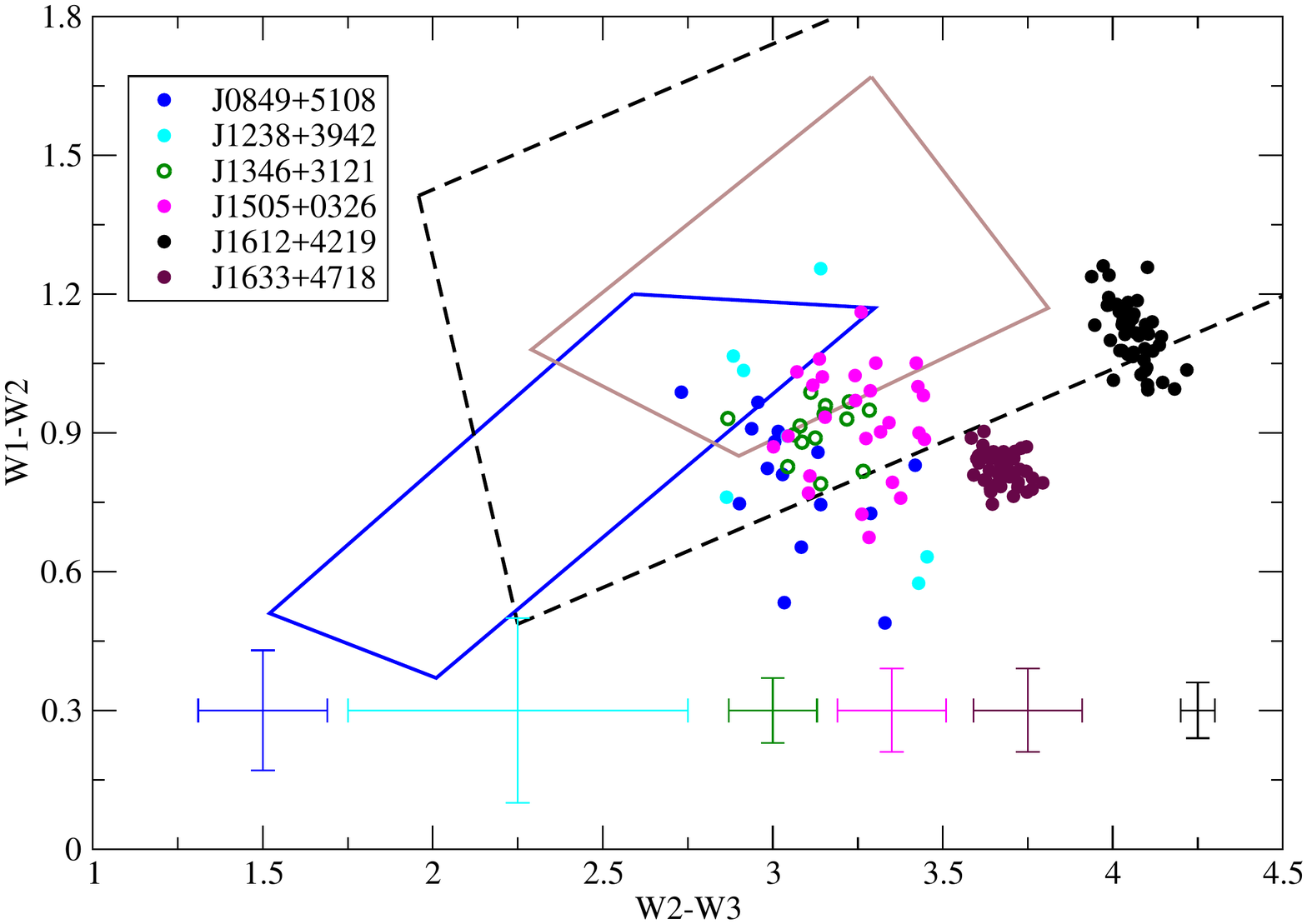}
\end{minipage}
\caption{Positions of the sample of NLS1 sources in the {\it WISE} color--color diagram. Solid lines represent the `WISE Gamma-ray Strip' for BL Lac objects (blue) and flat-spectrum quasars (brown) as defined by \cite{wgs1} and \cite{wgs2}. Dashed line shows the AGN wedge from \cite{AGN_wedge}. {\it Left panel:} Open circles represent the non-variable, filled circles the variable sources. Orange, green, and turquoise symbols stand for sources showing variability in both, only W1, and only W2 bands, respectively. The sizes of the filled symbols scale with the logarithm of the modulation index measured in band W1, except for those only variable in W2, for which they scale with the logarithm of the modulation index measured in band W2. The symbol sizes of the largest and smallest modulation indices are displayed in the lower right corner. Red symbols mark the sources detected in $\gamma$-rays by the {\it Fermi} satellite. The sources shown in the right panel are marked with circles of the same colors. {\it Right panel:} Simultaneous W1$-$W2 and W2$-$W3 colors of the $6$ variable sources which fall outside of the AGN wedge according to their photometric values given in the AllWISE catalog \cite{AllWISE}. Colored error bars at the bottom of the plot show the mean errors of the corresponding sources.}
\label{fig:w1w2w3}
\end{figure}

Most of the sources, $27$ out of the $42$ ($\sim 64$\%) showed variability in at least one of the IR bands of $3.4\,\mu$m and $4.6\,\mu$m; $20$ of them were variable in both bands, $7$ sources showed variability in only one of the bands, $5$ in W1 and $2$ in W2 only. All the $8$ sources in the list that have been detected in $\gamma$-rays by the {\it Fermi} satellite (reported in \cite{FermiLAT3rd}, \cite{FermiAGN1st}, \cite{J1644gamma}, and \cite{Foschini_gamma}) showed IR variability. Most of them both in band W1 and band W2. Two sources, J1102$+$2239 and J1246$+$0238, showed variability in one band only, in W2 and W1, respectively. These two sources were the faintest $\gamma$-ray emitter NLS1 objects among those discovered by \cite{Foschini_gamma}. 

In the other two bands, only one source showed possible variability, J0948$+$0022 had a $\chi^2_\mathrm{r}$ larger than 3 in band W3.

In the left panel of Fig. \ref{fig:w1w2w3}, we plot the {\it WISE} color--color diagram of the $42$ radio-loud NLS1 sources also indicating their variability behaviour. The colors were calculated using the photometric values given in the AllWISE catalog \cite{AllWISE}. Filled symbols are the variable objects, the symbol sizes are scaled with the logarithm of modulation indices. We overplotted the AGN wedge (defined by \cite{AGN_wedge}, dashed line) and the `WISE Gamma-ray strip' (WGS, defined by \cite{wgs1} and \cite{wgs2}) for flat-spectrum quasars (brown line) and for BL Lac objects (blue line). The fraction of variable sources in those which fall outside of the WGS is similar ($60$\%) to that of the whole sample. 

To investigate whether variability can explain the positions of these sources in the color--color plot, we calculated their colors using simultaneous measurements from the original WISE mission data taken in 2011. The mean errors for each plotted sources are shown at the bottom of the plot. In all sources but J1633$+$4718 significant change in at least one of the colors can be seen. J1238$+$3942 is the faintest among the sources in band W3 causing it to have the largest uncertainty in the W2$-$W3 color. It can be seen from the right panel of Fig. \ref{fig:w1w2w3} that variability can change the positions of sources significantly on the color--color diagram and such sources can easily wander in (and out of) the WGS or even the AGN wedge. The best example is J0849$+$5108, which is horizontally offset from all sources in the original WISE color--color plot (left panel), in sharp contrast with its locations in the right panel (blue dots) when using simultaneous measurements. 
There are two sources which remain within a well-defined small region in the color--color plot, J1612$+$4219 and J1633$+$4718. In both of them, the {\it WISE} measured IR emission can be influenced by star formation. J1633$+$4718 is an AGN and a starburst galaxy separated by $4''$ \cite{J1633pair}. These two objects cannot be resolved by {\it WISE}. According to the emission line study of Shirazi \& Brinchmann \cite{sf_agn}, both J1612$+$4219 and J1633$+$4718 are composite objects, in which the ionizing radiation originates from star formation and AGN.

Most of the variable sources, $18$ out of $27$, showed also variability at least in one band, when we checked each mission phase separately. We also detected short time scale variability in the sources (J0849$+$5180 and J0948$+$0022) previously reported by Jiang et al. \cite{jiang_rapid_wise} having intraday IR variability. We also detected short time scale variations in the third source, J1505$+$0526, studied by Jiang et al. \cite{jiang_rapid_wise}, occurring in a later mission phase after the publication of \cite{jiang_rapid_wise}. The most pronounced short time scale flux density changes are produced by J0948$+$0021, showing variations in every major epochs at both bands with modulation indices as high as $40$\%. All the $\gamma$-ray emitters except for the two faintest in $\gamma$-rays (J1102$+$2239 and J1246$+$0238) showed intraday IR variability. The light curve of J1644$+$2619, one of the $\gamma$-ray detected NLS1 is shown on Fig. \ref{fig:J1644}. The blazar nature of these sources is also supported by various very long baseline interferometric observations in the radio (e.g. \cite{mojave}).

\section{Summary}

Using the {\it WISE} mission data we investigated IR variability in a sample of radio-loud NLS1 sources. More than $60$\% of the sources showed variability at $3.4$ or $4.6\,\mu$m. Color variations can significantly influence the location of some sources in the {\it WISE} color--color diagnostic plot and may therefore influence their classification.
 
Intraday variability in at least one of the IR bands of $3.4$ or $4.6\,\mu$m can be detected in $66$\% of the variable sources. As described by Jiang et al. \cite{jiang_rapid_wise}, such short variability time scales imply a small size for the emitting region which is thus hard to reconcile with variations in the dusty torus. More probable explanation is that the IR emission is non-thermal and it originates from the jet. This scenario is further supported by the fact that the known bright $\gamma$-ray emitter NLS1 sources in the sample all showed intraday IR variability.



\bibliographystyle{JHEP}
\bibliography{ref}

\providecommand{\href}[2]{#2}\begingroup\raggedright\begin{thebibliography}{10}

\bibitem{Mathur2000}
S.~{Mathur}, \emph{{Narrow-line Seyfert 1 galaxies and the evolution of
  galaxies and active galaxies}},
  \href{https://doi.org/10.1046/j.1365-8711.2000.03530.x}{\emph{MNRAS}
  {\bfseries 314} (2000) L17}
  [\href{https://arxiv.org/abs/astro-ph/0003111}{{\ttfamily
  astro-ph/0003111}}].

\bibitem{CollinKawaguchi}
S.~{Collin} and T.~{Kawaguchi}, \emph{{Super-Eddington accretion rates in
  Narrow Line Seyfert 1 galaxies}},
  \href{https://doi.org/10.1051/0004-6361:20040528}{\emph{A\&A} {\bfseries 426}
  (2004) 797} [\href{https://arxiv.org/abs/astro-ph/0407181}{{\ttfamily
  astro-ph/0407181}}].

\bibitem{Komossa2006}
S.~{Komossa}, W.~{Voges}, D.~{Xu}, S.~{Mathur}, H.-M. {Adorf}, G.~{Lemson}
  et~al., \emph{{Radio-loud Narrow-Line Type 1 Quasars}},
  \href{https://doi.org/10.1086/505043}{\emph{AJ} {\bfseries 132} (2006) 531}
  [\href{https://arxiv.org/abs/astro-ph/0603680}{{\ttfamily
  astro-ph/0603680}}].

\bibitem{cacci}
A.~{Caccianiga}, S.~{Ant{\'o}n}, L.~{Ballo}, L.~{Foschini}, T.~{Maccacaro},
  R.~{Della Ceca} et~al., \emph{{WISE colours and star formation in the host
  galaxies of radio-loud narrow-line Seyfert 1}},
  \href{https://doi.org/10.1093/mnras/stv939}{\emph{MNRAS} {\bfseries 451}
  (2015) 1795} [\href{https://arxiv.org/abs/1504.07068}{{\ttfamily
  1504.07068}}].

\bibitem{wise}
E.~L. {Wright}, P.~R.~M. {Eisenhardt}, A.~K. {Mainzer}, M.~E. {Ressler}, R.~M.
  {Cutri}, T.~{Jarrett} et~al., \emph{{The Wide-field Infrared Survey Explorer
  (WISE): Mission Description and Initial On-orbit Performance}},
  \href{https://doi.org/10.1088/0004-6256/140/6/1868}{\emph{AJ} {\bfseries 140}
  (2010) 1868} [\href{https://arxiv.org/abs/1008.0031}{{\ttfamily 1008.0031}}].

\bibitem{neowise}
A.~{Mainzer}, J.~{Bauer}, R.~M. {Cutri}, T.~{Grav}, J.~{Masiero}, R.~{Beck}
  et~al., \emph{{Initial Performance of the NEOWISE Reactivation Mission}},
  \href{https://doi.org/10.1088/0004-637X/792/1/30}{\emph{ApJ} {\bfseries 792}
  (2014) 30} [\href{https://arxiv.org/abs/1406.6025}{{\ttfamily 1406.6025}}].

\bibitem{Stetson}
P.~B. {Stetson}, \emph{{On the Automatic Determination of Light-Curve
  Parameters for Cepheid Variables}},
  \href{https://doi.org/10.1086/133808}{\emph{PASP} {\bfseries 108} (1996)
  851}.

\bibitem{wgs1}
R.~{D'Abrusco}, F.~{Massaro}, M.~{Ajello}, J.~E. {Grindlay}, H.~A. {Smith} and
  G.~{Tosti}, \emph{{Infrared Colors of the Gamma-Ray-detected Blazars}},
  \href{https://doi.org/10.1088/0004-637X/748/1/68}{\emph{ApJ} {\bfseries 748}
  (2012) 68} [\href{https://arxiv.org/abs/1203.0568}{{\ttfamily 1203.0568}}].

\bibitem{wgs2}
F.~{Massaro}, R.~{D'Abrusco}, G.~{Tosti}, M.~{Ajello}, D.~{Gasparrini}, J.~E.
  {Grindlay} et~al., \emph{{The WISE Gamma-Ray Strip Parameterization: The
  Nature of the Gamma-Ray Active Galactic Nuclei of Uncertain Type}},
  \href{https://doi.org/10.1088/0004-637X/750/2/138}{\emph{ApJ} {\bfseries 750}
  (2012) 138} [\href{https://arxiv.org/abs/1203.1330}{{\ttfamily 1203.1330}}].

\bibitem{AGN_wedge}
S.~{Mateos}, A.~{Alonso-Herrero}, F.~J. {Carrera}, A.~{Blain}, M.~G. {Watson},
  X.~{Barcons} et~al., \emph{{Using the Bright Ultrahard XMM-Newton survey to
  define an IR selection of luminous AGN based on WISE colours}},
  \href{https://doi.org/10.1111/j.1365-2966.2012.21843.x}{\emph{MNRAS}
  {\bfseries 426} (2012) 3271}
  [\href{https://arxiv.org/abs/1208.2530}{{\ttfamily 1208.2530}}].

\bibitem{AllWISE}
R.~M. {Cutri} and {et al.}, \emph{{VizieR Online Data Catalog: WISE All-Sky
  Data Release (Cutri+ 2012)}}, {\emph{VizieR Online Data Catalog} {\bfseries
  2311} (2012) }.

\bibitem{FermiLAT3rd}
F.~{Acero}, M.~{Ackermann}, M.~{Ajello}, A.~{Albert}, W.~B. {Atwood},
  M.~{Axelsson} et~al., \emph{{Fermi Large Area Telescope Third Source
  Catalog}}, \href{https://doi.org/10.1088/0067-0049/218/2/23}{\emph{The
  Astrophysical Journal Supplement Series} {\bfseries 218} (2015) 23}
  [\href{https://arxiv.org/abs/1501.02003}{{\ttfamily 1501.02003}}].

\bibitem{FermiAGN1st}
A.~A. {Abdo}, M.~{Ackermann}, M.~{Ajello}, A.~{Allafort}, E.~{Antolini}, W.~B.
  {Atwood} et~al., \emph{{The First Catalog of Active Galactic Nuclei Detected
  by the Fermi Large Area Telescope}},
  \href{https://doi.org/10.1088/0004-637X/715/1/429}{\emph{ApJ} {\bfseries 715}
  (2010) 429} [\href{https://arxiv.org/abs/1002.0150}{{\ttfamily 1002.0150}}].

\bibitem{J1644gamma}
F.~{D'Ammando}, M.~{Orienti}, J.~{Larsson} and M.~{Giroletti}, \emph{{The first
  {$\gamma$}-ray detection of the narrow-line Seyfert 1 FBQS J1644+2619}},
  \href{https://doi.org/10.1093/mnras/stv1278}{\emph{MNRAS} {\bfseries 452}
  (2015) 520} [\href{https://arxiv.org/abs/1503.08226}{{\ttfamily
  1503.08226}}].

\bibitem{Foschini_gamma}
L.~{Foschini}, \emph{{Evidence of powerful relativistic jets in narrow-line
  Seyfert 1 galaxies}},  in \emph{Narrow-Line Seyfert 1 Galaxies and their
  Place in the Universe}, p.~24, 2011,
  \href{https://arxiv.org/abs/1105.0772}{{\ttfamily 1105.0772}}.

\bibitem{J1633pair}
W.~{Yuan}, B.~F. {Liu}, H.~{Zhou} and T.~G. {Wang}, \emph{{X-ray Observational
  Signature of a Black Hole Accretion Disk in an Active Galactic Nucleus RX
  J1633+4718}}, \href{https://doi.org/10.1088/0004-637X/723/1/508}{\emph{ApJ}
  {\bfseries 723} (2010) 508}
  [\href{https://arxiv.org/abs/1009.2808}{{\ttfamily 1009.2808}}].

\bibitem{sf_agn}
M.~{Shirazi} and J.~{Brinchmann}, \emph{{Strongly star forming galaxies in the
  local Universe with nebular He II{$\lambda$}4686 emission}},
  \href{https://doi.org/10.1111/j.1365-2966.2012.20439.x}{\emph{MNRAS}
  {\bfseries 421} (2012) 1043}
  [\href{https://arxiv.org/abs/1201.1290}{{\ttfamily 1201.1290}}].

\bibitem{jiang_rapid_wise}
N.~{Jiang}, H.-Y. {Zhou}, L.~C. {Ho}, W.~{Yuan}, T.-G. {Wang}, X.-B. {Dong}
  et~al., \emph{{Rapid Infrared Variability of Three Radio-loud Narrow-line
  Seyfert 1 Galaxies: A View from the Wide-field Infrared Survey Explorer}},
  \href{https://doi.org/10.1088/2041-8205/759/2/L31}{\emph{ApJL} {\bfseries
  759} (2012) L31} [\href{https://arxiv.org/abs/1210.2800}{{\ttfamily
  1210.2800}}].

\bibitem{mojave}
M.~L. {Lister}, M.~F. {Aller}, H.~D. {Aller}, D.~C. {Homan}, K.~I.
  {Kellermann}, Y.~Y. {Kovalev} et~al., \emph{{MOJAVE: XIII. Parsec-scale AGN
  Jet Kinematics Analysis Based on 19 years of VLBA Observations at 15 GHz}},
  \href{https://doi.org/10.3847/0004-6256/152/1/12}{\emph{AJ} {\bfseries 152}
  (2016) 12} [\href{https://arxiv.org/abs/1603.03882}{{\ttfamily 1603.03882}}].

\end{thebibliography}\endgroup



\end{document}